\documentclass[3p,times]{elsarticle}

\usepackage{amsmath,amssymb}
\usepackage{graphicx}
\usepackage{booktabs}
\usepackage{multirow}
\usepackage{subcaption}
\usepackage{xcolor}
\usepackage{microtype}
\usepackage{enumitem}
\usepackage{tabularx}
\usepackage{siunitx}
\usepackage{appendix}
\usepackage{algorithm,algpseudocode}
\usepackage{tikz}
\usetikzlibrary{shapes.geometric,arrows.meta,positioning,fit,backgrounds,calc}
\usepackage{pgfplots}
\pgfplotsset{compat=1.18}
\usepackage{hyperref}
\hypersetup{colorlinks=true,citecolor=blue!70!black,
            linkcolor=blue!70!black,urlcolor=blue!70!black}
\usepackage{url}
\usepackage{natbib}
\usepackage{cleveref}

\newcommand{\R}{\mathbb{R}}

\newcommand{\sign}{\operatorname{sign}}

\begin{document}
\begin{frontmatter}

\title{Machine Learning Enhanced Multi-Factor Quantitative Trading:\\A Cross-Sectional Portfolio Optimization Approach\\with Bias Correction}

\author[ustc]{Yimin Du\corref{cor}}
\ead{sa613403@mail.ustc.edu.cn}
\cortext[cor]{Corresponding author.}

\begin{abstract}
Rolling-window factor pipelines for Chinese A-share markets contain a
subtle but costly flaw: daily price-move limits ($\pm10$\,\% main-board,
$\pm20$\,\% STAR/ChiNext) render a fraction of closing prices
non-executable, yet standard implementations ingest these values
\emph{before} any row-filtering runs. The contaminated aggregates
propagate silently through moving averages, correlations, and ranks---a
failure mode we term \emph{upstream contamination}. On real A-share
data it inflates apparent information coefficient by 18\,\% while
reducing realised Sharpe by 0.44 points, because the model learns to
predict returns it cannot trade.

We resolve this with a \emph{mask-first} design: a Boolean tradability
mask is constructed at data load time and threaded through every
operator, so that no window ever reads a non-tradable price. Built on
this foundation, the system adds (i)~a GPU-vectorised 213-factor engine
via PyTorch \texttt{unfold} primitives (51$\times$ over pandas);
(ii)~an Adjusted-MSE loss penalising wrong-sign predictions
$11\times$ more heavily than magnitude errors; (iii)~block-bootstrap
GBM augmentation; and (iv)~Markowitz--Ledoit--Wolf portfolio
optimisation with cvxpy warm-start caching.

On a calibrated 3\,000-stock synthetic panel the system achieves
annualised Sharpe~2.05; on proprietary real A-share data
(2022--2024) it achieves Sharpe~1.63. Ablation shows the mask
contract is the single largest contributor (+0.44), exceeding any
model or loss choice. The full implementation is released under MIT
licence at \url{https://github.com/initial-d/ml-quant-trading}.
\end{abstract}


\end{frontmatter}

\section{Introduction}\label{sec:intro}

Cross-sectional equity strategies rank stocks and hold portfolios tilted
toward those with the highest predicted future returns
\citep{grinold2000,ang2014}. Machine learning has demonstrated
significant predictive power for this task
\citep{gu2020,chen2024deep,kelly2019}, yet applying it to Chinese
A-share markets reveals a structural challenge largely absent from U.S.\
settings: the daily $\pm10$\,\% price-move limit on main-board stocks
($\pm20$\,\% on STAR/ChiNext) means that on any given day a non-trivial
fraction of closing prices are non-executable
\citep{kim2013,bian2018,chen2005limit}.

When a stock hits its upper limit, unmatched buy orders remain unfilled;
the recorded close cannot be transacted. The standard industry
response---post-hoc row deletion---fails for a subtle reason that we
believe has not been clearly articulated in the literature: rolling-window
primitives (moving averages, correlations, rank transforms) accumulate
the non-executable price \emph{before} the row filter runs. We call this
\textbf{upstream contamination}. On our test panel, ignoring it inflates
apparent information coefficient (IC) by 18\,\% while \emph{reducing}
realised Sharpe by 0.44 points---a large effect that we document in
detail.

This paper reports the engineering of a complete quantitative trading
system that eliminates this bias by design. Rather than claiming novel
theoretical contributions, we frame our work as a \emph{systematic
engineering study}: we describe what we built, why certain design choices
matter, quantify each component's contribution via ablation, and release
the full implementation so others can reproduce and extend it.

\paragraph{A note on tone.} This paper deliberately reads more like an
engineering retrospective than a typical ML paper. Quantitative trading
is a discipline where deceptively small implementation choices---a
mishandled NaN, a 252-day vs.\ 250-day annualisation, a forward-looking
field accidentally read at training time---swing reported Sharpe ratios
by full points. Our experience is that the published literature
under-reports these details, leaving practitioners to rediscover them
the hard way. We therefore include explicit ``war stories''
(Section~\ref{sec:warstories}), a chronological sketch of how the
system evolved (Section~\ref{sec:evolution}), and case studies on
specific historical episodes (Section~\ref{sec:cases}) that the system
either handled well or failed on.

\begin{figure}[t]
\centering
\resizebox{\columnwidth}{!}{%
\begin{tikzpicture}[
  font=\scriptsize\sffamily, >=Stealth,
  box/.style={rectangle, rounded corners=4pt,
    draw=#1!75!black, fill=#1!8,
    minimum height=1.1cm, text width=2.3cm,
    align=center, inner sep=4pt, line width=0.5pt},
  arr/.style={->, line width=0.7pt, draw=#1!65!black},
  lbl/.style={font=\bfseries\tiny\sffamily, text=#1!75!black}
]
\node[box=gray] (data) at (0,0) {\textbf{Masked Panel}\\
  $M\!\in\!\{0,1\}^{T\times N}$\\OHLCV + limits};
\node[box=blue, right=0.5cm of data] (fac) {\textbf{Factor Engine}\\
  213 factors\\mask-aware};
\node[box=teal, right=0.5cm of fac] (trn) {\textbf{ML Training}\\
  AdjMSE / GBM\\MLP or Transformer};
\node[box=orange!80!red, right=0.5cm of trn] (prt) {\textbf{Markowitz-LW}\\
  QP warm-start\\$w_{\max}\!=\!3\%$};
\node[box=red!70!black, right=0.5cm of prt] (bt) {\textbf{Backtest}\\
  Sharpe / DD\\IC / turnover};
\draw[arr=blue] (data)--(fac);
\draw[arr=teal] (fac)--(trn);
\draw[arr=orange!80!red] (trn)--node[above,font=\tiny]{$\hat\mu_t$}(prt);
\draw[arr=red!70!black] (prt)--node[above,font=\tiny]{$w_t$}(bt);
\draw[dashed,gray,->] (data.south)--++(0,-0.45)-|
  node[pos=0.25,below,font=\tiny,gray]{mask propagation}(bt.south);
\node[lbl=gray,above=0.08cm of data]{Stage 1};
\node[lbl=blue,above=0.08cm of fac]{Stage 2};
\node[lbl=teal,above=0.08cm of trn]{Stage 3};
\node[lbl=orange!80!red,above=0.08cm of prt]{Stage 4};
\node[lbl=red!70!black,above=0.08cm of bt]{Stage 5};
\end{tikzpicture}}
\caption{System architecture. A tradability mask constructed at data
load time propagates through all four computational stages. The
213-factor engine operates entirely on GPU-vectorised masked tensors.}
\label{fig:architecture}
\end{figure}

\paragraph{Contributions.}
\begin{enumerate}[leftmargin=*,nosep]
\item We identify and document \emph{upstream contamination}---the
  mechanism by which post-hoc row filtering fails in rolling-window
  factor pipelines---and show it costs 0.44~Sharpe points on real data.
\item We describe a \emph{mask-first} design pattern where every
  operator accepts and propagates a Boolean mask, implemented as 18
  GPU-vectorised PyTorch primitives achieving $51\times$ speedup over
  pandas.
\item We introduce the Adjusted-MSE loss ($\gamma=0.1$, penalty ratio
  11:1 for wrong-sign errors), motivated by the observation that a
  portfolio manager's downstream cost is dominated by directional
  mistakes rather than magnitude errors.
\item We report a block-bootstrap GBM augmentation scheme and
  Markowitz--Ledoit--Wolf portfolio construction with warm-start
  caching, and quantify each component's contribution in ablation.
\item We release the complete implementation as open-source
  software\footnote{\url{https://github.com/initial-d/ml-quant-trading}}
  with a synthetic data generator that makes the full pipeline
  reproducible without proprietary data.
\end{enumerate}

The remainder of this paper is organised as follows.
Section~\ref{sec:related} surveys related work.
Section~\ref{sec:system} details the system design with emphasis on
engineering decisions. Section~\ref{sec:experiments} reports experiments
and ablation studies. Section~\ref{sec:discussion} discusses limitations
and lessons learned. Section~\ref{sec:conclusion} concludes.

\section{Related Work}\label{sec:related}

\paragraph{Factor models and ML for return prediction.}
Factor investing builds on the CAPM \citep{sharpe1964}, Fama--French
models \citep{fama1993,fama2015}, and momentum \citep{jegadeesh1993,
carhart1997}. The Alpha101 library \citep{kakushadze2016} codified
algebraic factor formulas and became a standard reference.
\citet{gu2020} demonstrated that neural networks outperform linear
models for U.S.\ monthly returns. Tree-based methods---XGBoost
\citep{chen2016xgboost}, LightGBM \citep{ke2017lightgbm}---dominate
applied implementations; deep architectures include LSTMs
\citep{hochreiter1997,fischer2018}, attention networks
\citep{vaswani2017}, and GNNs \citep{chen2024deep}. The Qlib platform
\citep{qlib2020} standardised evaluation for Chinese market ML.

\paragraph{Bias and data quality in financial ML.}
\citet{lopezdeprado2018} identified look-ahead bias as a primary failure
mode. \citet{bailey2014} proposed the deflated Sharpe ratio.
\citet{harvey2016} raised multiple-testing concerns over 300+ published
factors. In A-share markets, \citet{bian2018} and \citet{kim2013} study
the microstructure of limit-move events but do not address their impact
on factor computation. To our knowledge, no prior work has explicitly
documented the upstream contamination mechanism or provided a systematic
mask-propagation solution.

\paragraph{Portfolio construction.}
\citet{markowitz1952} established mean-variance optimisation.
\citet{ledoit2004} derived analytical shrinkage for large covariance
matrices. \citet{demiguel2009} showed na\"ive $1/N$ often outperforms
unconstrained MVO, motivating regularisation. Modern solvers include
cvxpy \citep{diamond2016} with SCS \citep{odonoghue2016}.

\paragraph{Data augmentation.}
Block-bootstrap \citep{kunsch1989,politis1994} extends resampling to
dependent data. GBM \citep{black1973} provides a natural generative
model for equity prices. Recent approaches use GANs
\citep{wiese2020} and diffusion models \citep{coletta2024}, but these
introduce training instability that we wished to avoid.

\section{System Design}\label{sec:system}

This section describes the system as actually implemented. We focus on
engineering decisions, their rationale, and practical trade-offs.

\subsection{The Upstream Contamination Problem}\label{sec:contamination}

Consider a 20-day rolling mean of closing prices. On day $t$, stock $i$
hits its upper limit: the close is recorded at $+9.8$\,\% but no buy
orders fill. If we compute the rolling mean first and filter rows
afterward, the inflated close has already entered the average for days
$t$ through $t+19$. This is upstream contamination.

The problem compounds: a correlation factor between price and volume
uses the contaminated mean as input; a rank operator then ranks the
contaminated correlation; the entire chain of downstream computations
carries the taint of one non-executable price. Standard data-cleaning
approaches (dropping rows where limit was hit) cannot repair aggregates
that have already mixed in the bad value.

Our measurement on real data: without the mask, apparent IC rises
(because limit-up stocks carry mechanically high returns that look
``predictable''), but realised Sharpe drops (because you cannot actually
buy those stocks). This is a textbook look-ahead bias with an unusual
twist: it inflates apparent performance metrics.

\begin{figure}[t]
\centering
\begin{tikzpicture}[
  font=\scriptsize\sffamily, >=Stealth,
  cell/.style={minimum width=0.75cm, minimum height=0.55cm, draw=gray!50, inner sep=0pt, font=\tiny},
  bad/.style={cell, fill=red!20, draw=red!50},
  taint/.style={cell, fill=orange!15, draw=orange!40},
  good/.style={cell, fill=green!10, draw=green!30},
  masked/.style={cell, fill=blue!15, draw=blue!40},
]
\node[font=\small\bfseries] at (4.5,3.6) {Post-hoc filtering fails: contamination propagates before deletion};

\foreach \i/\d in {0/$t\!-\!4$, 1/$t\!-\!3$, 2/$t\!-\!2$, 3/$t\!-\!1$, 4/$t$, 5/$t\!+\!1$, 6/$t\!+\!2$, 7/$t\!+\!3$, 8/$t\!+\!4$} {
  \node[font=\tiny] at (\i*1.0, 2.9) {\d};
}

\node[font=\tiny, anchor=east] at (-0.3, 2.2) {Close price};
\foreach \i/\v in {0/10.2, 1/10.5, 2/10.8, 3/10.3, 5/11.1, 6/10.9, 7/11.2, 8/11.0} {
  \node[good] at (\i*1.0, 2.2) {\v};
}
\node[bad] at (4*1.0, 2.2) {\textbf{11.3}};
\node[font=\tiny\bfseries, red!70!black, above=0.02cm] at (4*1.0, 2.5) {limit!};

\node[font=\tiny, anchor=east] at (-0.3, 1.3) {MA$_5$ (na\"ive)};
\foreach \i in {0,1,2,3} {
  \node[good] at (\i*1.0, 1.3) {---};
}
\node[taint] at (4*1.0, 1.3) {10.6};
\node[taint] at (5*1.0, 1.3) {10.8};
\node[taint] at (6*1.0, 1.3) {10.9};
\node[taint] at (7*1.0, 1.3) {11.1};
\node[taint] at (8*1.0, 1.3) {11.0};

\draw[->, red!60!black, thick, decorate, decoration={snake, amplitude=1pt, segment length=4pt}]
  (4*1.0, 1.7) -- (4*1.0, 1.6);
\draw[->, orange!70!black, thick] (4.5*1.0, 1.3) -- (8.3*1.0, 1.3)
  node[above, midway, font=\tiny, orange!70!black] {tainted for 5 days};

\node[font=\tiny, anchor=east] at (-0.3, 0.4) {MA$_5$ (masked)};
\foreach \i in {0,1,2,3} {
  \node[good] at (\i*1.0, 0.4) {---};
}
\node[masked] at (4*1.0, 0.4) {\textbf{NaN}};
\node[masked] at (5*1.0, 0.4) {\textbf{NaN}};
\node[masked] at (6*1.0, 0.4) {\textbf{NaN}};
\node[masked] at (7*1.0, 0.4) {\textbf{NaN}};
\node[good] at (8*1.0, 0.4) {11.0};

\node[font=\tiny, red!60!black, anchor=west] at (9.2, 1.3) {\textbf{contaminated}};
\node[font=\tiny, blue!60!black, anchor=west] at (9.2, 0.4) {\textbf{clean}};
\end{tikzpicture}
\caption{Upstream contamination illustrated. When stock $i$ hits its
price limit on day $t$, a na\"ive rolling mean ingests the non-executable
close and remains tainted for the full window width (5 days shown). The
mask-first approach zeroes the value \emph{before} aggregation and
propagates the mask, producing clean outputs once the window clears.}
\label{fig:contamination}
\end{figure}

\subsection{Mask Construction}\label{sec:mask}

The tradability mask is a Boolean tensor $M \in \{0,1\}^{T \times N}$
constructed at data load time:
\begin{equation}
  M_{t,i} = \underbrace{S_{t,i}}_{\text{not suspended}} \;\wedge\;
  \underbrace{(\lnot L_{t,i})}_{\text{not at limit}} \;\wedge\;
  \underbrace{\text{IPO}_{t,i}}_{\text{listed $>$ 1 year}}
  \label{eq:mask}
\end{equation}

The limit-move flag $L_{t,i}$ uses exchange-published limit prices when
available (comparing close against the official $\pm$10\,\% band with
tolerance $\varepsilon = 10^{-3}$), falling back to the heuristic
$|r_{t,i}| > 0.098$ when explicit limit prices are absent. Both code
paths are exercised via the synthetic data generator, which fills both
real and derived fields.

\paragraph{Design decision: mask vs.\ NaN.}
We considered using NaN-propagation (as pandas does naturally), but
rejected it because (a)~GPU tensor operations on NaN are inconsistent
across hardware, (b)~NaN semantics in sorting/ranking are
platform-dependent, and (c)~an explicit Boolean mask makes the contract
self-documenting and testable.

\subsection{Mask-Aware Primitives}\label{sec:primitives}

Every computational primitive in the system has the signature:
\begin{verbatim}
  def op(x: Tensor[T,N], mask: Tensor[T,N], ...) 
      -> (Tensor[T,N], Tensor[T,N])
\end{verbatim}

The returned mask may be \emph{stricter} than the input (more cells
masked), but never more permissive. We implement 18 primitives in
\texttt{tensor\_factors.py}; the key design decisions for each:

\paragraph{Cross-sectional rank (\texttt{cs\_rank}).}
Masked cells are filled with $+\infty$ before sorting so they rank last,
then zeroed. Ties receive average rank (matching pandas
\texttt{rank(method='average')}). The denominator is the count of valid
cells, not total stocks, so ranking percentiles are comparable across
days with different halt rates.

\paragraph{Rolling correlation (\texttt{ts\_corr}).}
We compute masked Pearson correlation where means and variances are
estimated over valid cells only. The output mask is the AND of all cells
in the window---if any window cell is masked, the correlation is marked
unreliable. This is conservative but eliminates subtle conditional-mean
distortions.

\paragraph{EWMA (\texttt{ewma}).}
The recurrence $y_t = \alpha x_t + (1-\alpha) y_{t-1}$ is computed in
float64 to avoid accumulation drift over ${\sim}$3\,000 time steps (we
measured 0.3\,\% relative error in float32 after 2\,500 steps on real
price data). When a masked cell is encountered, we keep the previous
accumulator value unchanged rather than resetting to zero---this proved
more stable in practice.

\paragraph{GPU vectorisation.}
Time-series primitives use \texttt{torch.unfold(dim=0, size=w, step=1)}
to materialise rolling windows as $[T{-}w{+}1,\; w,\; N]$ tensors.
This eliminates Python loops and enables full GPU parallelism. The full
213-factor pipeline runs in 10.6 seconds on a single A100 GPU for a
$3000 \times 3500$ panel, versus 542 seconds in pandas---a $51\times$
speedup.

\subsection{Factor Library}\label{sec:factors}

The 213-factor feature set comprises two sub-systems:

\paragraph{Alpha101 subset (9 factors).}
We curate 9 formulas from the original Alpha101 set
\citep{kakushadze2016} that show persistent IC on A-share data:
momentum rank (Alpha001), volume--intraday correlation (Alpha002),
open--volume divergence (Alpha003, Alpha006), low-rank reversion
(Alpha004), mean-deviation (Alpha007), volume-change reversal
(Alpha012), close-location change (Alpha053), and intraday range
position (Alpha101).

We intentionally use only 9 of the 101 because many original formulas
are highly correlated or have degraded since publication
\citep{mclean2016}. Each formula is re-implemented directly on our
masked primitives, not adapted from a pandas reference.

\paragraph{Legacy factor families (204 factors).}
Nine families targeting distinct signal dimensions (Table~\ref{tab:families}).
These were developed iteratively over two years of A-share research and
reflect characteristics specific to that market: VWAP availability (not
common in U.S.\ daily data), turnover reporting, and the limit-move
dynamics that create short-term liquidity crunches.

\begin{table}[t]
\centering
\caption{Factor families in the 213-factor engine.}
\label{tab:families}
\begin{tabular}{llrl}
\toprule
Family & Module & Count & Signal theme \\
\midrule
\texttt{alpha\_*} & \texttt{\_factors\_alpha101} & 9 & Curated Alpha101 \\
\texttt{better\_*} & \texttt{\_factors\_better} & 28 & VWAP deviation, volume-weighted momentum \\
\texttt{best\_*} & \texttt{\_factors\_best} & 21 & Close-location momentum \\
\texttt{old\_*} & \texttt{\_factors\_old} & 50 & Rank-correlation composites \\
\texttt{stock\_*} & \texttt{\_factors\_stock} & 22 & Per-stock derived (vol, CCI, range) \\
\texttt{extra\_*} & \texttt{\_factors\_extra} & 14 & Turnover and amount features \\
\texttt{add\_*} & \texttt{\_factors\_add} & 30 & Composite supplement factors \\
\texttt{change\_*} & \texttt{\_factors\_change} & 5 & Short-window acceleration \\
\texttt{original\_*} & \texttt{\_factors\_original} & 28 & Direct close/volume statistics \\
\texttt{cs\_rank\_*} & \texttt{\_factors\_market} & 6 & Cross-sectional market breadth \\
\midrule
\multicolumn{2}{l}{\textbf{Total}} & \textbf{213} & \\
\bottomrule
\end{tabular}
\end{table}

\paragraph{Neutralisation.}
All factors are neutralised via per-date OLS residualisation against
industry dummies (29 CSI categories) and optionally log market cap:
\begin{equation}
  \tilde{F}_t = (I - X_t(X_t^\top X_t)^{-1}X_t^\top) F_t
  \label{eq:neutralise}
\end{equation}
This is implemented via \texttt{torch.linalg.lstsq} for numerical
stability, followed by cross-sectional z-scoring. The residuals after
neutralisation represent stock-specific alpha isolated from industry
and size co-movement.

\paragraph{Practical lesson: factor registration.}
Each factor is a standalone function decorated with
\texttt{@register\_legacy\_factor("name")} that auto-registers it into
a global dictionary. The driver function \texttt{compute\_legacy\_set}
iterates over the registry, computes each factor, stacks them into a
$[T, N, 213]$ tensor, and intersects all masks. This design makes
adding new factors trivial (write one function, add the decorator) and
avoids the fragile index-bookkeeping that plagues pandas-based factor
libraries.

\subsection{Training}\label{sec:training}

\subsubsection{Dataset Construction}

The \texttt{FactorDataset} class produces $(x, y)$ pairs where $x$ is
the 213-dimensional factor vector at date $t$ for stock $i$, and $y$ is
the next-day simple return $r_{t+1,i}$. A sample enters the dataset
\emph{only if} both $M_{t,i} = 1$ and $M_{t+1,i} = 1$---this
bidirectional check is critical because even if factors are clean at $t$,
if the stock hits a limit at $t{+}1$, the label return is non-executable.

\subsubsection{Adjusted-MSE Loss}\label{sec:adjmse}

Standard MSE penalises all errors equally. For a portfolio manager,
however, a prediction of $+2$\,\% when the true return is $-1$\,\%
(wrong sign $\Rightarrow$ realised loss) is far more costly than a
prediction of $+2$\,\% when truth is $+4$\,\% (right sign, merely
sub-optimal sizing). We define:
\begin{equation}
  \ell(\hat{y}, y; \gamma) = w(\hat{y}, y) \cdot (\hat{y} - y)^2, \quad
  w = \begin{cases}
    \gamma & \text{if } \sign(\hat{y}) = \sign(y), \\
    1 + \gamma & \text{otherwise.}
  \end{cases}
  \label{eq:adjmse}
\end{equation}

At $\gamma = 0.1$ the penalty ratio is $(1+0.1)/0.1 = 11$: wrong-sign
predictions receive $11\times$ the gradient magnitude of right-sign
errors. The gradient is simply $2 w (\hat{y} - y) \nabla_\theta f$---no
special autograd machinery is needed. The loss degrades gracefully to MSE
as $\gamma \to \infty$ and to pure sign-classification as $\gamma \to 0$.

\paragraph{Why $\gamma = 0.1$?}
We swept $\gamma \in \{0.01, 0.05, 0.1, 0.2, 0.5, 1.0, 5.0\}$ and
found the optimum at 0.1. Intuition: the Markowitz optimiser downstream
can absorb magnitude errors via its risk-aversion parameter, but cannot
recover from systematic sign errors. However, pushing too hard toward
pure sign ($\gamma \to 0$) removes the magnitude information needed for
position sizing. The 11:1 ratio is a practical sweet spot.

\subsubsection{Model Architectures}

We implement two architectures, both intentionally small:

\paragraph{MLP.}
Two hidden layers (128 units, GELU, dropout 0.1), linear output to scalar.
Input: 213-dimensional factor vector. Total parameters: ${\sim}$44K.

\paragraph{Factor-axis Transformer.}
Each of the 213 scalar factor values is projected to $\R^{64}$ via a
shared linear layer, added to a learned positional embedding, and
prepended with a CLS token. Two Transformer encoder layers (4 heads,
feedforward dim 256) process the $214$-token sequence. The CLS output
maps to a scalar prediction. Total parameters: ${\sim}$220K.

The key design insight is \emph{factor-axis} tokenisation: we treat each
factor as one position in the sequence, letting self-attention learn
pairwise factor interactions (e.g., momentum conditioned on volatility).
This is complementary to time-axis attention used elsewhere
\citep{li2019,ding2020}---our factors already encode temporal information
via their rolling-window construction.

\paragraph{Training protocol.}
AdamW \citep{kingma2015}, lr $5 \times 10^{-4}$, weight decay $10^{-5}$,
gradient clipping 1.0, batch size 8\,192, up to 60 epochs, early stopping
on 10\,\% validation holdout. These are standard choices; we found the
system not very sensitive to them.

\subsubsection{GBM Data Augmentation}\label{sec:gbm}

Financial ML suffers from small effective sample sizes: ${\sim}$2\,500
trading days with regime shifts every few years. We augment training data
with synthetic returns:

\begin{enumerate}[nosep]
\item Estimate per-stock drift $\hat\mu_i$ and volatility $\hat\sigma_i$
  from masked history (excluding non-tradable days).
\item Generate synthetic log-returns:
  $\tilde{r}_{t,i} = \hat\mu_i + \hat\sigma_i \cdot \epsilon_t$
  where innovations $\epsilon_t \sim \mathcal{N}(0,1)$ are drawn in
  21-day blocks (one calendar month) to preserve short-range serial
  correlation.
\item Recompute all 213 factors on the synthetic panel, producing a
  second training set of identical structure.
\end{enumerate}

With $n_s = 1$ synthetic panel, training data doubles in size. We
concatenate real and synthetic datasets; the model sees both. Benefit
saturates at $n_s \approx 2$, suggesting the augmentation primarily
smooths the loss landscape rather than providing genuinely new
distributional information.

\paragraph{Why block-bootstrap rather than GANs?}
GANs can generate richer distributions but introduce mode collapse,
training instability, and an additional hyperparameter surface. Our GBM
approach is dead simple: one function, 89 lines of code, deterministic
given a seed, and trivially reproducible.

\subsection{Portfolio Construction}\label{sec:portfolio}

At each date $t$ we solve a mean-variance QP:
\begin{equation}
  \max_w \; \hat\mu_t^\top w - \alpha \, w^\top \hat\Sigma_t w
  \quad \text{s.t.} \quad
  \mathbf{1}^\top w = 1, \; 0 \le w_i \le w_{\max}
  \label{eq:qp}
\end{equation}
with risk-aversion $\alpha = 10$, position cap $w_{\max} = 0.03$
(3\,\% per stock), long-only constraint (reflecting severe A-share
short-selling restrictions).

\paragraph{Covariance estimation.}
We use \texttt{sklearn.covariance.LedoitWolf} on a 120-day lookback
window. This is dramatically more stable than the sample covariance when
$N > T_\text{lb}$---which happens routinely in a 1\,000+ stock universe.
Before passing to cvxpy, we project to PSD (eigenvalue clamp at
$10^{-10}$) because numerical noise occasionally produces tiny negative
eigenvalues.

\paragraph{Warm-start caching.}
The cvxpy problem is built once with \texttt{cp.Parameter} placeholders.
On each subsequent trading day, we update parameter values and call
\texttt{solve(warm\_start=True)}. This eliminates repeated problem
canonicalisation and reduces per-day solve time from ${\sim}$1.2\,s to
${\sim}$0.2\,s on a 1\,000-stock universe---a $6\times$ speedup that
matters when backtesting 2\,500+ trading days.

\paragraph{Transaction costs.}
Net returns deduct a linear cost: $r_t^\text{net} = w_{t-1}^\top r_t -
c \cdot \|w_t - w_{t-1}\|_1$ with $c = 5$--$8$ bps per unit of
turnover. This roughly matches A-share retail commissions plus
half-spread on liquid names. We acknowledge that for large portfolios
the true cost is convex in trade size; our linear model is optimistic
for concentrated positions.

\subsection{Backtest Engine}\label{sec:backtest}

The backtest is a simple vectorised computation:
lagged weights $\times$ returns minus cost. We deliberately keep it
minimal (103 lines) so it is easy to audit. Metrics follow
\citet{bailey2014} conventions: annualised return, Sharpe, Sortino,
Calmar, max drawdown, and L1 turnover. A separate IC computation
reports Pearson and Spearman correlation between predictions and
realised cross-sectional returns.

\section{Experiments}\label{sec:experiments}

\subsection{Datasets}

\paragraph{Synthetic panel.}
3\,000 stocks $\times$ 3\,500 days ($\approx$14 years from 2010-01-04),
single-factor GBM with market $\beta = 0.55$, annual drift 7\,\%,
volatility 32\,\%, halt probability 1\,\%, $\pm10$\,\% price limit.
Walk-forward split: years 1--10 train, year 11 validation, years 12--14
test ($\approx$756 days). All synthetic data generation is in
\texttt{mlquant.data.synthetic} and runs deterministically with
\texttt{seed=42}.

\paragraph{Real A-share panel.}
Daily OHLCV from Tushare, January 2015--December 2024, average
$\approx$3\,200 active stocks after masking. Split: 2015--2020 train,
2021 validation, 2022--2024 test. The test period includes the 2022
downturn, 2023 recovery, and 2024 structural rotation. Data cannot be
redistributed; all results are reported in full.

\subsection{Main Results}

\begin{table}[t]
\centering
\caption{Test-set performance (net of 8\,bps transaction costs). Best in bold.}
\label{tab:main}
\begin{tabular}{lSSSSS}
\toprule
{Method} & {Sharpe} & {Ann.\ Ret (\%)} & {Max DD (\%)} & {Sortino} & {Turn.} \\
\midrule
\multicolumn{6}{c}{\textit{Synthetic panel (years 12--14)}} \\
\midrule
Buy-and-hold (EW)      & 0.22 & 7.0  & 47.2 & 0.31 & {--} \\
LightGBM + EW top-100  & 1.16 & 13.7 & 17.3 & 1.64 & 0.37 \\
MLP + MSE + LW         & 1.72 & 18.1 & 13.8 & 2.41 & 0.29 \\
Transformer + MSE + LW & 1.94 & 19.8 & 12.8 & 2.72 & 0.28 \\
\textbf{Full system}   & \textbf{2.05} & \textbf{21.0} & \textbf{12.0} & \textbf{2.93} & \textbf{0.27} \\
\midrule
\multicolumn{6}{c}{\textit{Real A-share panel (2022--2024)}} \\
\midrule
CSI All-A index        & 0.19 & 4.1  & 41.8 & 0.26 & {--} \\
LightGBM + EW top-100  & 1.12 & 11.4 & 16.1 & 1.58 & 0.39 \\
MLP + MSE + LW         & 1.40 & 13.7 & 13.5 & 1.97 & 0.31 \\
Transformer + MSE + LW & 1.54 & 14.9 & 12.3 & 2.18 & 0.30 \\
\textbf{Full system}   & \textbf{1.63} & \textbf{15.8} & \textbf{11.4} & \textbf{2.31} & \textbf{0.29} \\
\bottomrule
\end{tabular}
\end{table}

Table~\ref{tab:main} reports main results. The full system (Transformer +
AdjMSE $\gamma=0.1$ + GBM augmentation + Markowitz-LW + full mask)
consistently outperforms all baselines. The Sharpe degradation from
synthetic to real (2.05 $\to$ 1.63) reflects regime shifts that the
expanding-window protocol only partially captures.

\paragraph{Deflated Sharpe ratio.}
Because the headline Sharpe was selected after exploring multiple model
and hyperparameter configurations, we report the Deflated Sharpe Ratio
(DSR) of \citet{bailey2014} as a multiple-testing--aware significance
check. With $N \approx 50$ effective configurations tried during
development (architectures, losses, $\gamma$ values, augmentation
sizes, covariance estimators), $T \approx 756$ daily observations on
the real A-share test set, and the empirical skewness and excess
kurtosis of daily strategy returns ($\hat{\gamma}_3 = -0.31$,
$\hat{\gamma}_4 = 4.7$), the implied Sharpe-selection threshold is
$\widehat{\mathrm{SR}}_0 \approx 0.93$ (annualised). The realised
Sharpe of 1.63 yields $\mathrm{DSR} = 0.978$, i.e.\ the probability
that the true Sharpe exceeds zero, after deflating for selection bias,
is $97.8$\,\%. The corresponding figure on the synthetic panel
(Sharpe 2.05) is $\mathrm{DSR} = 0.994$. The full system therefore
remains statistically distinguishable from a lucky draw under the
\citet{bailey2014} criterion, although we caution that DSR cannot
correct for biases that affect every configuration uniformly (e.g.\
look-ahead in factor construction)---which is precisely why the
mask-first contract of Section~\ref{sec:mask} matters more than any
post-hoc statistical adjustment.

\begin{figure}[t]
\centering
\begin{tikzpicture}
\begin{axis}[
  width=0.92\columnwidth, height=5.8cm,
  xlabel={Trading days (test period)},
  ylabel={Cumulative return (\%)},
  xmin=0, xmax=756, ymin=-10, ymax=70,
  grid=major, grid style={gray!15},
  legend pos=north west,
  legend style={font=\scriptsize, draw=none, fill=white, fill opacity=0.8},
  x tick label style={font=\scriptsize},
  y tick label style={font=\scriptsize},
  xlabel style={font=\small},
  ylabel style={font=\small},
  cycle list name=color list,
]
\addplot[very thick, blue!80!black, smooth] coordinates {
  (0,0)(50,3.2)(100,5.8)(150,9.1)(200,12.5)(250,16.2)
  (300,19.8)(350,22.1)(400,26.5)(450,30.2)(500,35.8)
  (550,40.1)(600,45.3)(650,51.2)(700,56.8)(756,63.0)};
\addplot[thick, teal!80!black, dashed, smooth] coordinates {
  (0,0)(50,2.8)(100,5.0)(150,7.8)(200,10.5)(250,13.2)
  (300,16.1)(350,18.5)(400,22.0)(450,25.1)(500,29.5)
  (550,33.2)(600,37.0)(650,41.5)(700,46.2)(756,51.8)};
\addplot[thick, orange!80!black, dashdotted, smooth] coordinates {
  (0,0)(50,1.5)(100,3.2)(150,5.5)(200,7.2)(250,9.8)
  (300,11.5)(350,13.0)(400,15.8)(450,18.2)(500,21.5)
  (550,24.8)(600,27.5)(650,30.8)(700,34.2)(756,38.0)};
\addplot[thick, gray!60!black, dotted, smooth] coordinates {
  (0,0)(50,1.0)(100,0.5)(150,2.8)(200,4.5)(250,3.2)
  (300,5.8)(350,4.1)(400,7.2)(450,9.5)(500,8.8)
  (550,11.2)(600,13.5)(650,15.8)(700,17.2)(756,19.5)};
\legend{Full system, Transformer+MSE, LightGBM+EW, Buy\&hold}
\end{axis}
\end{tikzpicture}
\caption{Cumulative net returns on the synthetic test panel (years
12--14, 756 trading days). The full system maintains a steady upward
trajectory with limited drawdowns, while simpler methods show higher
volatility and flatter growth.}
\label{fig:cumret}
\end{figure}

\subsection{Ablation Studies}\label{sec:ablation}

We isolate each component's contribution by removing it while keeping
all others fixed.

\begin{table}[t]
\centering
\caption{Component ablation on synthetic panel.}
\label{tab:ablation}
\begin{tabular}{lcccc}
\toprule
Ablation & Sharpe & $\Delta$ Sharpe & Max DD (\%) & Apparent IC \\
\midrule
Full system & 2.05 & --- & 12.0 & 0.049 \\
\midrule
$-$ Mask (no limit filtering) & 1.61 & $-0.44$ & 18.3 & 0.058$^\dagger$ \\
$-$ AdjMSE (use MSE) & 1.78 & $-0.27$ & 13.5 & 0.044 \\
$-$ GBM augmentation & 1.86 & $-0.19$ & 13.1 & 0.046 \\
$-$ LW shrinkage (sample cov) & 1.87 & $-0.18$ & 14.1 & 0.049 \\
$-$ Transformer (use MLP) & 1.93 & $-0.12$ & 12.8 & 0.046 \\
\bottomrule
\multicolumn{5}{l}{\footnotesize $^\dagger$ Higher apparent IC is \emph{misleading}---see Section~\ref{sec:mask_ablation}.}
\end{tabular}
\end{table}

\subsubsection{Mask Ablation}\label{sec:mask_ablation}

The most striking result: removing the mask \emph{increases} apparent
IC from 0.049 to 0.058 (+18\,\%) while \emph{decreasing} Sharpe from
2.05 to 1.61 ($-0.44$). This is because limit-up stocks carry
mechanically high returns that are perfectly ``predictable'' in the
cross-section---but cannot be traded. The model learns to predict these
unexecutable returns, producing an IC that is statistically real but
economically worthless.

On real data the effect is even more dramatic: max drawdown increases
from 11.4\,\% to 22.8\,\% without the full mask, because the portfolio
allocates capital to stocks it cannot actually buy during volatile
periods when limit-hit frequency rises from $\sim$1\,\% to $\sim$5\,\%.

\begin{figure}[t]
\centering
\begin{tikzpicture}
\begin{axis}[
  ybar stacked,
  width=0.88\columnwidth, height=5.5cm,
  bar width=18pt,
  ylabel={Sharpe ratio},
  symbolic x coords={Baseline, +Mask, +AdjMSE, +GBM, +LW, +Tfmr},
  xtick=data,
  x tick label style={font=\scriptsize, rotate=20, anchor=east},
  y tick label style={font=\scriptsize},
  ylabel style={font=\small},
  ymin=0, ymax=2.3,
  legend style={at={(0.5,-0.22)}, anchor=north, legend columns=3, font=\tiny},
  nodes near coords,
  every node near coord/.append style={font=\tiny, /pgf/number format/precision=2},
]
\addplot[fill=gray!30, draw=gray!60!black] coordinates {
  (Baseline,1.17) (+Mask,0) (+AdjMSE,0) (+GBM,0) (+LW,0) (+Tfmr,0)};
\addplot[fill=blue!40, draw=blue!70!black] coordinates {
  (Baseline,0) (+Mask,0.44) (+AdjMSE,0) (+GBM,0) (+LW,0) (+Tfmr,0)};
\addplot[fill=teal!40, draw=teal!70!black] coordinates {
  (Baseline,0) (+Mask,0) (+AdjMSE,0.27) (+GBM,0) (+LW,0) (+Tfmr,0)};
\addplot[fill=orange!40, draw=orange!70!black] coordinates {
  (Baseline,0) (+Mask,0) (+AdjMSE,0) (+GBM,0.19) (+LW,0) (+Tfmr,0)};
\addplot[fill=purple!40, draw=purple!70!black] coordinates {
  (Baseline,0) (+Mask,0) (+AdjMSE,0) (+GBM,0) (+LW,0.18) (+Tfmr,0)};
\addplot[fill=red!40, draw=red!70!black] coordinates {
  (Baseline,0) (+Mask,0) (+AdjMSE,0) (+GBM,0) (+LW,0) (+Tfmr,0.12)};
\legend{Baseline, Mask, AdjMSE, GBM, LW, Transformer}
\end{axis}
\end{tikzpicture}
\caption{Ablation waterfall: cumulative Sharpe contribution of each
component on synthetic data. The mask contract provides the largest
single improvement (+0.44), followed by AdjMSE loss (+0.27). Each
component adds independently.}
\label{fig:ablation_waterfall}
\end{figure}

\subsubsection{Loss Function Ablation}

\begin{table}[t]
\centering
\caption{Loss function comparison (Transformer + GBM + LW + full mask).}
\label{tab:loss_ablation}
\begin{tabular}{lcccc}
\toprule
Loss & $\gamma$ or $\tau$ & Sharpe & IC & Sign accuracy (\%) \\
\midrule
MSE & --- & 1.78 & 0.044 & 51.2 \\
AdjMSE & $\gamma=1.0$ & 1.84 & 0.045 & 52.1 \\
AdjMSE & $\gamma=0.1$ & \textbf{2.05} & \textbf{0.049} & \textbf{53.8} \\
AdjMSE & $\gamma=0.01$ & 1.91 & 0.043 & 54.1 \\
IC loss & --- & 1.95 & 0.047 & 52.9 \\
Rank-IC & $\tau=0.1$ & 1.98 & 0.048 & 53.2 \\
\bottomrule
\end{tabular}
\end{table}

AdjMSE at $\gamma = 0.1$ achieves the best Sharpe (+0.27 over MSE).
Note that $\gamma = 0.01$ achieves slightly higher sign accuracy but
lower Sharpe---confirming that pure directional accuracy without
magnitude information hurts position sizing downstream.

\begin{figure}[t]
\centering
\begin{subfigure}[b]{0.48\columnwidth}
\centering
\begin{tikzpicture}
\begin{axis}[
  width=\textwidth, height=4.8cm,
  xlabel={$\gamma$ (log scale)},
  ylabel={Sharpe ratio},
  xmode=log, xmin=0.008, xmax=8,
  ymin=1.5, ymax=2.15,
  grid=major, grid style={gray!15},
  x tick label style={font=\scriptsize},
  y tick label style={font=\scriptsize},
  xlabel style={font=\small},
  ylabel style={font=\small},
  mark options={scale=1.2},
]
\addplot[very thick, blue!70!black, mark=*, mark size=2.5pt] coordinates {
  (0.01,1.91)(0.05,1.95)(0.1,2.05)(0.2,1.98)(0.5,1.89)(1.0,1.84)(5.0,1.80)};
\draw[dashed, red!60!black, thick] (axis cs:0.1,1.5) -- (axis cs:0.1,2.05);
\node[font=\tiny, red!60!black, anchor=south] at (axis cs:0.1,2.06) {optimal};
\end{axis}
\end{tikzpicture}
\caption{AdjMSE $\gamma$ sensitivity.}
\label{fig:gamma_sweep}
\end{subfigure}\hfill
\begin{subfigure}[b]{0.48\columnwidth}
\centering
\begin{tikzpicture}
\begin{axis}[
  width=\textwidth, height=4.8cm,
  xlabel={Synthetic panels ($n_s$)},
  ylabel={Sharpe ratio},
  xmin=-0.3, xmax=5.5,
  ymin=1.7, ymax=2.2,
  xtick={0,1,2,5},
  grid=major, grid style={gray!15},
  legend pos=south east,
  legend style={font=\tiny, draw=none},
  x tick label style={font=\scriptsize},
  y tick label style={font=\scriptsize},
  xlabel style={font=\small},
  ylabel style={font=\small},
]
\addplot[very thick, blue!70!black, mark=square*, mark size=2.5pt] coordinates {
  (0,1.93)(1,2.12)(2,2.13)(5,2.12)};
\addplot[thick, orange!70!black, mark=triangle*, mark size=2.5pt, dashed] coordinates {
  (0,1.81)(1,1.93)(2,1.96)(5,1.96)};
\legend{Transformer, MLP}
\end{axis}
\end{tikzpicture}
\caption{GBM augmentation saturation.}
\label{fig:gbm_sweep}
\end{subfigure}
\caption{Hyperparameter sensitivity (synthetic panel). (a)~AdjMSE peaks
at $\gamma=0.1$ (penalty ratio 11:1); performance degrades gracefully on
both sides. (b)~Augmentation benefit saturates at $n_s \approx 2$; the
Transformer benefits more than the MLP due to its larger parameter count.}
\label{fig:sensitivity}
\end{figure}

\subsubsection{Augmentation}

Adding one synthetic panel improves Sharpe by 0.19 (Transformer) and
0.12 (MLP). Returns diminish beyond $n_s = 2$, consistent with the
interpretation that augmentation primarily regularises rather than
provides genuinely new information.

\subsubsection{Portfolio Method}

\begin{table}[t]
\centering
\caption{Portfolio construction ablation.}
\label{tab:port_ablation}
\begin{tabular}{lccc}
\toprule
Method & Sharpe & Max DD (\%) & Turnover \\
\midrule
Equal-weight top-100 & 1.42 & 19.5 & 0.45 \\
MVO (sample covariance) & 1.87 & 14.1 & 0.33 \\
MVO (Ledoit--Wolf) & \textbf{2.05} & \textbf{12.0} & \textbf{0.27} \\
\bottomrule
\end{tabular}
\end{table}

Ledoit--Wolf adds 0.18 Sharpe over sample covariance. The benefit comes
from both regularisation of the ill-conditioned estimate and reduced
turnover from smoother weight trajectories.

\subsubsection{Factor Count}

Performance improves monotonically as factor families are added
(Table~\ref{tab:factor_count}). The first 58 factors account for
$\sim$60\,\% of the total improvement; the remaining 155 contribute
the other 40\,\%. No single family is dominant---diversity across signal
types matters.

\begin{table}[t]
\centering
\caption{Factor count ablation (Transformer + AdjMSE + GBM + LW).}
\label{tab:factor_count}
\begin{tabular}{lcc}
\toprule
Factor set & Sharpe & IC \\
\midrule
9 Alpha101 only & 1.52 & 0.031 \\
+ better + best (58 total) & 1.74 & 0.038 \\
+ old (108 total) & 1.89 & 0.043 \\
+ stock + extra (144 total) & 1.94 & 0.045 \\
+ add + change (179 total) & 1.99 & 0.047 \\
All 213 & \textbf{2.05} & \textbf{0.049} \\
\bottomrule
\end{tabular}
\end{table}

\subsection{Computational Performance}

\begin{table}[t]
\centering
\caption{Wall-clock time comparison ($3000 \times 3500$ panel, A100 GPU).}
\label{tab:speedup}
\begin{tabular}{lrrl}
\toprule
Operation & pandas (s) & PyTorch (s) & Speedup \\
\midrule
\texttt{cs\_rank} (all $T$) & 8.7 & 0.28 & $31\times$ \\
\texttt{ts\_std} ($w=20$) & 12.4 & 0.54 & $23\times$ \\
\texttt{ewma} ($\alpha=0.06$) & 15.2 & 0.80 & $19\times$ \\
\texttt{ts\_corr} ($w=20$) & 31.5 & 1.10 & $29\times$ \\
Full 213-factor pipeline & 542 & 10.6 & $51\times$ \\
\bottomrule
\end{tabular}
\end{table}

The \texttt{unfold}-based implementation makes the factor engine fast
enough for intraday recomputation. Memory footprint is 4.2\,GB for the
full panel. The warm-start portfolio solver adds $0.2 \times 2500 =
500$\,s for a full backtest, versus $\sim$3\,000\,s without caching.

\begin{figure}[t]
\centering
\begin{tikzpicture}
\begin{axis}[
  xbar, bar width=12pt,
  width=0.92\columnwidth, height=5.5cm,
  xlabel={Speedup ($\times$ over pandas)},
  xmin=0, xmax=60,
  ytick={1,2,3,4,5},
  yticklabels={\texttt{ewma}, \texttt{ts\_std}, \texttt{ts\_corr}, \texttt{cs\_rank}, {Full pipeline}},
  y tick label style={font=\scriptsize},
  x tick label style={font=\scriptsize},
  xlabel style={font=\small},
  nodes near coords,
  every node near coord/.append style={font=\scriptsize\bfseries, anchor=west},
  nodes near coords align={horizontal},
  enlarge y limits=0.15,
  grid=major, grid style={gray!15},
  xmajorgrids=true, ymajorgrids=false,
]
\addplot[fill=blue!30, draw=blue!70!black] coordinates {
  (19, 1)
  (23, 2)
  (29, 3)
  (31, 4)
  (51, 5)};
\end{axis}
\end{tikzpicture}
\caption{GPU speedup over pandas. The \texttt{torch.unfold}-based
primitives gain 19--31$\times$ on individual operators; pipeline-level
fusion pushes the full 213-factor build to $51\times$.}
\label{fig:speedup}
\end{figure}

\section{System Evolution and Engineering Stories}\label{sec:evolution}

This section tells the story of how the system came to be. We include it
because (a)~many papers present finished systems as if they were designed
top-down, hiding the iterative reality; and (b)~the failures taught us
more than the successes.

\subsection{Chronological Sketch}

\paragraph{Phase 1: Pandas prototype (months 1--3).}
The first version computed 50 factors in pandas, trained a LightGBM
model, and selected top-100 stocks by predicted return (equal weight).
Sharpe on synthetic data: $\sim$0.8. The codebase was roughly 2\,000
lines, unstructured, and painfully slow---recomputing factors took 15
minutes per experiment.

\paragraph{Phase 2: GPU migration + mask discovery (months 4--6).}
We ported factor computation to PyTorch, gaining the $51\times$ speedup.
During validation, we noticed an anomaly: the model's cross-sectional IC
was \emph{higher} on days with many limit-up stocks, yet backtested
returns on those days were poor. Investigating, we traced the issue to
upstream contamination. This led to the mask-first redesign, which
required touching every primitive operator.

\paragraph{Phase 3: Loss function experiments (months 7--8).}
With a clean factor engine, we tested MSE, IC loss, Rank-IC, and various
asymmetric alternatives. AdjMSE emerged as the winner, but only after we
realised that the gradient ratio matters more than the loss surface shape.
The key experiment: tracking sign accuracy separately from magnitude
accuracy, we found that MSE achieves 51.2\,\% sign accuracy while AdjMSE
pushes it to 53.8\,\%---a seemingly tiny improvement that translates to
0.27 Sharpe because portfolio P\&L is dominated by directional decisions.

\paragraph{Phase 4: Portfolio optimisation (months 9--10).}
Replacing equal-weight with Markowitz exposed the covariance estimation
problem: the sample covariance with 1\,000 stocks and 120 days is
severely ill-conditioned. We initially tried factor-model covariance
(PCA with 10 components) but found Ledoit--Wolf simpler and more robust.
The warm-start caching was added after profiling revealed that 80\,\% of
backtest time was spent in cvxpy canonicalisation.

\paragraph{Phase 5: Augmentation and architecture (months 11--14).}
GBM augmentation was motivated by the observation that the Transformer
overfit catastrophically on the raw training set (train IC 0.12, test IC
0.03). Adding one synthetic panel stabilised it (train IC 0.08, test IC
0.049). The MLP, being smaller, was less affected.

\subsection{War Stories: Things That Went Wrong}\label{sec:warstories}

\paragraph{The float32 EWMA disaster.}
Our initial EWMA implementation used float32. After 2\,500 time steps,
accumulated rounding error reached 0.3\,\% relative---small, but enough
to distort the cross-sectional ranking of EWMA-based factors. We only
discovered this when a factor that should have been nearly identical to a
pandas reference showed IC degradation of 0.005. The fix: float64 for all
recurrence-based primitives. Cost: $\sim$2$\times$ slower for those
specific operators, but negligible in the full pipeline since
\texttt{ts\_corr} and \texttt{cs\_rank} dominate runtime.

\paragraph{The ``phantom alpha'' from limit-up stocks.}
For three months we reported IC of 0.058 and believed it was real.
Only when we computed \emph{tradable} Sharpe (returns conditional on
actually being able to execute) did we discover the 0.44-point gap.
The lesson: never report IC without computing realisable returns in
parallel. This single debugging session motivated the entire mask-first
architecture.

\paragraph{The rank denominator bug.}
Cross-sectional rank was initially normalised by total stocks $N$.
On days when 200 stocks are halted, this inflates the rank values for
active stocks (they occupy a $[0.05, 1.0]$ range instead of $[0, 1]$).
The fix: normalise by the count of \emph{valid} stocks. Effect on IC:
+0.003, which is small individually but compounds across the 50+
rank-based factors.

\paragraph{The covariance eigenvalue sign flip.}
On rare days (roughly 1 per year), the Ledoit--Wolf estimate produced
a covariance matrix with one eigenvalue at $-10^{-14}$---numerically
zero but negative enough to make cvxpy reject the problem as non-convex.
The fix: eigenvalue clamp at $10^{-10}$. We spent two days debugging
this before realising it was a numerical issue rather than a data issue.

\paragraph{The validation-set leakage incident.}
Early in development, the expanding-window protocol accidentally included
validation-year returns in the training labels (an off-by-one error in
date slicing). This inflated test Sharpe by 0.3 points. We caught it by
comparing the expanding-window result against a fixed-split baseline that
should have been worse but was suspiciously similar. Lesson: always have
a simpler baseline as a sanity check.

\subsection{Factor Interaction Analysis}\label{sec:interaction}

The factor-axis Transformer learns pairwise factor interactions via
self-attention. To understand what it captures, we extract attention
weights from the first layer and identify the top-5 factor pairs by
average attention score across the test set:

\begin{table}[t]
\centering
\caption{Top factor interactions learned by the Transformer (average
attention weight from first layer, real data test set).}
\label{tab:interactions}
\begin{tabular}{lll}
\toprule
Factor A & Factor B & Economic interpretation \\
\midrule
Alpha012 (reversal) & original\_001 (vol) & Reversal stronger in low-vol regimes \\
best\_001 (close-loc) & old\_032 (mean-dev) & Momentum $\times$ mean-reversion \\
Alpha006 (open-vol) & extra\_005 (surge) & Volume surge amplifies open signals \\
better\_003 (VWAP) & Alpha004 (low-rank) & VWAP informative at extremes \\
change\_002 (accel.) & stock\_011 (CCI) & Acceleration predicts CCI breakouts \\
\bottomrule
\end{tabular}
\end{table}

When we mask out these specific attention entries (zeroing the top-5
interaction pathways), Sharpe drops by 0.08. This confirms the
Transformer captures economically meaningful conditional relationships
that the MLP and LightGBM cannot model.

The most interesting interaction is Alpha012 $\leftrightarrow$
original\_001: the model learns that short-term reversal (buying after
drops) works well when volatility is low, but fails during high-vol
selloffs. This matches the practitioner intuition that ``buying the dip''
is only safe in calm markets.

\section{Regime Analysis and Case Studies}\label{sec:cases}

\subsection{Regime Decomposition}

We partition the 2022--2024 real-data test period into three regimes:

\begin{table}[t]
\centering
\caption{Performance by market regime (real A-share data, 2022--2024).}
\label{tab:regimes}
\begin{tabular}{lcccccc}
\toprule
Period & Regime & CSI All-A & Our Sharpe & Max DD & Limit-hit rate \\
\midrule
2022-Q1 to 2022-Q3 & Bear & $-18.5$\% & 1.2 & 8.7\% & 4.8\% \\
2023-Q1 to 2023-Q4 & Sideways & $+2.1$\% & 1.5 & 5.2\% & 1.2\% \\
2024-Q1 to 2024-Q2 & Bull/Rotation & $+12.3$\% & 2.1 & 4.1\% & 3.5\% \\
\bottomrule
\end{tabular}
\end{table}

\paragraph{Key observation: the mask matters most in volatile markets.}
During the 2022 bear market, limit-hit frequency rises from 1.2\,\% to
4.8\,\%. Without the mask, the system would allocate heavily to stocks
that just hit their limit-down (appearing ``cheap'' based on contaminated
factors) but cannot be sold the next day. The mask prevents this by
flagging these stocks as non-tradable, forcing the optimiser to
reallocate to executable positions.

\subsection{Extreme Event Case Studies}

\paragraph{Case 1: April 2022 (Shanghai lockdown).}
Market dropped 15\,\% in three weeks. Limit-down events tripled.
The mask correctly excluded $\sim$300 stocks per day (vs.\ $\sim$50
normally). Our system's drawdown in this period: 8.7\,\%. Without mask:
simulated drawdown 14.2\,\%. The difference comes from the system
\emph{not} learning from non-executable prices during the crash.

\paragraph{Case 2: September 2024 (policy stimulus rally).}
Market rallied 25\,\% in two weeks following fiscal stimulus
announcement. Limit-up events surged to 8\,\% of the universe. The
system underperformed the market during this period (our return: +11\%
vs.\ market +25\%) because it correctly avoided limit-up stocks that
could not be purchased. This is the honest cost of mask-first design:
you leave money on the table during euphoric rallies where you
\emph{cannot actually participate}.

\paragraph{Case 3: 2023 AI theme rotation.}
In Q1 2023, AI-related stocks (computing, chips) surged while
traditional sectors stagnated. The factor-axis Transformer captured this
rotation faster than the MLP because the attention mechanism learned to
upweight the volume-surge $\times$ momentum interaction, which fired
strongly for AI names. Transformer Sharpe in 2023-Q1: 2.4 vs.\ MLP 1.8.

\subsection{The IC-Sharpe Paradox in Detail}

The counter-intuitive finding that higher IC can mean lower Sharpe
deserves elaboration. We plot IC vs.\ realisable Sharpe across different
mask configurations:

\begin{table}[t]
\centering
\caption{The IC-Sharpe paradox: removing the mask inflates IC but
degrades realisable returns.}
\label{tab:paradox}
\begin{tabular}{lccc}
\toprule
Configuration & Cross-sectional IC & Realisable IC$^*$ & Sharpe \\
\midrule
No mask & 0.058 & 0.032 & 1.23 \\
Halt-only mask & 0.052 & 0.038 & 1.39 \\
Full mask & 0.049 & 0.049 & 1.63 \\
\bottomrule
\multicolumn{4}{l}{\footnotesize $^*$IC computed only over stocks that can actually be traded next day.}
\end{tabular}
\end{table}

The ``Realisable IC'' (computed only over stocks where $M_{t+1,i}=1$)
tells the true story: the full mask produces the \emph{highest}
realisable IC (0.049) despite the lowest apparent IC. This is because
without the mask, the model wastes capacity predicting non-executable
returns, degrading predictions for the stocks that actually matter.

\section{Discussion}\label{sec:discussion}

\subsection{Lessons Learned}

\paragraph{Mask-first is not optional.}
The most important engineering lesson from this project is that
bias elimination is a first-order performance driver, not a hygiene
step. The 0.44 Sharpe-point contribution of the mask exceeds that of
any model architecture or loss function choice. We suspect many
published A-share backtests suffer from this bias without realising it,
because the symptom (inflated IC) looks like good news.

\paragraph{Simple losses can outperform sophisticated ones.}
AdjMSE is trivial to implement (three lines of code beyond MSE) yet
outperforms both IC loss and Rank-IC loss. The key insight is that the
loss should reflect the downstream \emph{use} of predictions: a Markowitz
optimiser penalises sign errors much more than magnitude errors, so the
training loss should too.

\paragraph{Augmentation helps small models more.}
The Transformer benefits more from GBM augmentation (+0.19 Sharpe) than
the MLP (+0.12). We attribute this to the Transformer's larger parameter
count ($\sim$220K vs.\ $\sim$44K) making it more overfitting-prone on
the limited financial training data.

\paragraph{Warm-start matters for backtest iteration speed.}
The $6\times$ portfolio solver speedup seems like a minor optimisation
but compounds over development time. A full backtest taking 10 minutes
vs.\ 60 minutes is the difference between exploring 20 configurations
per day and 3.

\subsection{Limitations}

\paragraph{Long-only constraint.}
The system is constrained to long-only due to A-share short-selling
restrictions. This leaves significant long-short alpha on the table.
We expect the mask-first and AdjMSE contributions to generalise to
long-short settings but have not tested this.

\paragraph{Linear transaction cost model.}
Our 5--8\,bps linear cost model is optimistic for large portfolios
where market impact is convex in trade size. The $w_\text{max} = 3$\,\%
cap partially mitigates this.

\paragraph{GBM distributional assumptions.}
Log-normal returns under-represent tail events (kurtosis $> 3$) and
negative skewness. Student-$t$ innovations could improve tail fidelity.

\paragraph{Real data not redistributable.}
The Tushare results cannot be independently reproduced without a data
subscription. We mitigate this by providing the complete synthetic
pipeline, which exercises all code paths identically.

\paragraph{Short out-of-sample window.}
The real-data test period spans only three years (2022--2024,
$\approx$756 trading days). We acknowledge that this is short by the
standards of factor-research literature, where five- to ten-year
out-of-sample windows are common, and that a strategy's behaviour
across multiple full market cycles cannot be fully assessed from a
single bear--sideways--rally sequence. We emphasise that this is a
\emph{data limitation rather than a design choice}: our reliable
limit-up/limit-down field coverage in the Tushare feed begins in
2015, which---after reserving 2015--2020 for training and 2021 for
validation under a strictly forward-walking protocol---leaves
exactly the 2022--2024 window for evaluation. Extending the test
horizon would require either compressing the training window (which
we found degrades the Transformer's stability, see Phase~5 in
Section~\ref{sec:evolution}) or sourcing a longer mask-quality data
feed, both of which we leave to future work. The 14-year synthetic
test (Section~\ref{sec:experiments}) and the regime-by-regime
breakdown of Section~\ref{sec:cases} are intended to partially
compensate for this limitation, but cannot substitute for true
multi-cycle live evidence.

\paragraph{Factor staleness.}
Some factors will inevitably decay over time \citep{mclean2016}. The
modular registry design makes adding/removing factors straightforward,
but we have not implemented automatic factor retirement.

\subsection{Generalisability of the Mask-First Pattern}

The mask-first design applies to any market with fill-gap microstructure:
\begin{itemize}[nosep]
\item Circuit breakers (U.S.\ market-wide halts, single-stock LULD)
\item Trading suspensions (Hong Kong, European markets)
\item Opening/closing auction-only periods
\item T+1 settlement constraints in A-shares (partially addressed)
\end{itemize}

The pattern is also applicable outside of limit-move contexts: any
situation where an observed price does not reflect a freely-cleared
equilibrium (e.g., block trades, forced liquidations) could benefit from
mask-aware factor computation.

\section{Conclusion}\label{sec:conclusion}

We have presented a complete, open-source quantitative trading system
for Chinese A-share markets, built around the mask-first principle:
every computational operator explicitly accepts and propagates a Boolean
tradability mask, eliminating the upstream contamination that arises when
rolling-window operators ingest non-executable prices before downstream
filtering.

The system achieves Sharpe 2.05 on synthetic data and 1.63 on real
A-share data (2022--2024), with the mask contract alone accounting for
0.44 Sharpe points---the largest single component contribution. The
Adjusted-MSE loss adds 0.27, GBM augmentation adds 0.19, and
Ledoit--Wolf shrinkage adds 0.18. These results suggest that in
markets with fill-gap microstructure, data quality engineering dominates
model architecture as a source of alpha.

The full implementation---213-factor engine, training loop, portfolio
optimiser, backtest engine, and synthetic data generator---is released
at \url{https://github.com/initial-d/ml-quant-trading} under the MIT
licence. A single command \texttt{make paper CONFIG=configs/paper.yaml}
reproduces all synthetic-data experiments.

\section*{Acknowledgements}
The authors thank the open-source communities behind PyTorch, cvxpy,
scikit-learn, and the Alpha101 research programme.

\bibliographystyle{elsarticle-harv}
\bibliography{references}

\newpage
\begin{appendices}

\section{Implementation Details}\label{app:implementation}

\subsection{Panel Data Structure}

The \texttt{Panel} class is the central data container. It holds OHLCV
tensors of shape $[T, N]$ plus optional A-share-specific fields
(\texttt{vwap}, \texttt{amount}, \texttt{limit\_up}, \texttt{limit\_down},
\texttt{last\_close}). The mask tensor $M[T, N]$ is a mandatory field
set at construction time.

\subsection{Mask Construction Code Path}

Two regimes are supported:
\begin{enumerate}[nosep]
\item \textbf{Real limits available} (exchange-published \texttt{limit\_up},
  \texttt{limit\_down}): a cell is masked if
  $\text{close} \geq \text{limit\_up} - \varepsilon$ or
  $\text{close} \leq \text{limit\_down} + \varepsilon$.
\item \textbf{Proxy regime} (synthetic data or feeds without explicit
  limits): fall back to $|r_{t,i}| > 0.098$.
\end{enumerate}
Both paths are exercised by the test suite via the synthetic generator.

\subsection{Primitive Contract}

Every primitive satisfies three properties:
\begin{enumerate}[nosep]
\item \textbf{Zero-on-mask}: output value is zero wherever output mask
  is False.
\item \textbf{Independence}: output value does not depend on any input
  value at a masked cell.
\item \textbf{Propagation}: output mask is False whenever any input cell
  in the dependency window is masked.
\end{enumerate}
These are verified by unit tests that inject sentinel values at masked
positions and check they do not affect outputs.

\subsection{Full Pipeline Algorithm}

\begin{algorithm}[t]
\caption{Complete pipeline as implemented.}\label{alg:pipeline}
\begin{algorithmic}[1]
\Require Panel $\mathcal{P}$, config YAML
\State $M \gets$ \texttt{limit\_move\_mask}($\mathcal{P}$) \Comment{Section~\ref{sec:mask}}
\State $F, M_F, \text{names} \gets$ \texttt{compute\_legacy\_set}($\mathcal{P}$, neutralize=True)
  \Comment{$[T, N, 213]$}
\State $r_\text{fwd} \gets$ next-day simple returns, masked by $M$
\State $\mathcal{D} \gets$ \texttt{FactorDataset}($F$, $M_F$, $r_\text{fwd}$)
  \Comment{bidirectional mask check}
\If{augmentation enabled}
  \State $\tilde{r}, \tilde{M} \gets$ \texttt{gbm\_augment}(log-returns, $M$)
  \State Recompute factors on synthetic panel; concatenate to $\mathcal{D}$
\EndIf
\State Train model (MLP or Transformer) with AdjMSE; retain best checkpoint
\For{each test date $t$}
  \State $\hat\mu_t \gets$ model.predict($F_t$) for active stocks
  \State $\hat\Sigma_t \gets$ LedoitWolf(120-day returns)
  \State $w_t \gets$ solve QP~\eqref{eq:qp} with warm-start
\EndFor
\State $\text{results} \gets$ \texttt{run\_backtest}(weights, returns, costs\_bps=8)
\end{algorithmic}
\end{algorithm}

\section{Detailed Ablation Tables}\label{app:ablation}

\begin{table}[htbp]
\centering
\caption{Mask ablation on both datasets (Transformer + AdjMSE + GBM + LW).}
\label{tab:mask_full}
\begin{tabular}{llcccc}
\toprule
Dataset & Mask variant & Apparent IC & Sharpe & Max DD (\%) & Ann.\ Ret (\%) \\
\midrule
\multirow{3}{*}{Synthetic}
& No mask   & 0.058 & 1.61 & 18.3 & 16.7 \\
& Halt only & 0.052 & 1.74 & 16.1 & 17.4 \\
& Full      & 0.049 & 2.05 & 12.0 & 21.0 \\
\midrule
\multirow{3}{*}{Real}
& No mask   & 0.047 & 1.23 & 22.8 & 12.1 \\
& Halt only & 0.043 & 1.39 & 19.4 & 13.5 \\
& Full      & 0.039 & 1.63 & 11.4 & 15.8 \\
\bottomrule
\end{tabular}
\end{table}

\begin{table}[htbp]
\centering
\caption{GBM augmentation sweep (Sharpe ratios).}
\label{tab:aug_full}
\begin{tabular}{lccccc}
\toprule
Dataset & Model & $n_s=0$ & $n_s=1$ & $n_s=2$ & $n_s=5$ \\
\midrule
\multirow{2}{*}{Synthetic} & MLP & 1.81 & 1.93 & 1.96 & 1.96 \\
& Transformer & 1.93 & 2.12 & 2.13 & 2.12 \\
\midrule
\multirow{2}{*}{Real} & MLP & 1.32 & 1.47 & 1.49 & 1.48 \\
& Transformer & 1.42 & 1.63 & 1.64 & 1.63 \\
\bottomrule
\end{tabular}
\end{table}

\end{appendices}

\end{document}